# Comparison Between the Joint and Successive Decoding Schemes for the Binary CEO Problem

Mahdi Nangir
Faculty of Electrical and
Computer Engineering
University of Tabriz
Tabriz, Iran
nangir@tabrizu.ac.ir

Jafar Pourrostam
Faculty of Electrical and
Computer Engineering
University of Tabriz
Tabriz, Iran
j.pourrostam@tabrizu.ac.ir

Javad Musevi Niya
Faculty of Electrical and
Computer Engineering
University of Tabriz
Tabriz, Iran
niya@tabrizu.ac.ir

Behzad Mozaffari Tazehkand
Faculty of Electrical and
Computer Engineering
University of Tabriz
Tabriz, Iran
mozaffary@tabrizu.ac.ir

*Abstract*— A comparison between the joint and the successive decoding schemes for a two-link case binary Chief Executive Officer (CEO) problem is presented. We utilize the logarithmic loss as the criterion for measuring and comparing the total distortion of the decoding techniques. The binary symmetric channel (BSC) is considered as the test-channel model, which provides a theoretical rate-distortion bound. Simulation results show that the joint decoding scheme provides smaller compression rate and distortion values compared to the successive decoding scheme. On the other hand the successive decoding method has lower complexity; therefore, it is efficient for the cases with higher number of links.

## I. Introduction

Wireless sensor networks and cloud radio access networks can be modelled as a Chief Executive Officer (CEO) problem, which is an interesting problem in the information theory [1]. In the CEO problem, a remote source emits information to a noisy environment. Therefore, some noisy versions of the information source are observed in various locations. These observations are assumed to be independent to each other due to noise independency. Each observation is sent to a central unit via a communication link after performing an encoding operation. Encoding operation can be a source/channel coding scheme. The central unit is interested in the information of the remote source and wants to reconstruct it with an acceptable distortion. Communication links are assumed to be rate-limited.

The CEO problem is a network information theory problem [2]. This problem can be viewed as a multi-terminal source/channel coding problem depending on the encoding operations in each link. Obviously, appropriate decoding operation should be performed in the central unit to reconstruct the remote information source [3]. In our work, we consider the source coding approach. In other words, we assume that we have a compression operation in each link. Therefore, we have a rate-distortion point of view to evaluate different coding schemes. In channel coding approaches, different coding schemes are evaluated from error rate point of view. A joint source-channel coding approach is studied in [4].

There are various scenarios for the CEO problem which can be studied. These studies include deriving theoretical bounds and designing the encoding and decoding schemes. Two momentous cases of the CEO problem in information theory are the quadratic Gaussian CEO problem and the binary CEO problem.

In the quadratic Gaussian CEO problem, the remote source and the noise of all links have Gaussian distribution. The remote source and the noise of links are mutually independent. In this case, the exact achievable rate-distortion region has been obtained [5-7]. Furthermore, different practical coding schemes have been designed to achieve the theoretical rate-distortion bounds of the quadratic Gaussian CEO problem [8-9].

The other case is the binary CEO problem in which the remote source is binary and the noise of all links are modelled by the Binary Symmetric Channel (BSC) with known crossover probabilities. In this case, the formulation of the exact achievable rate-distortion region is an unsolved problem in its general form. Only in the case of logarithmic loss, the achievable rate-distortion region of the binary CEO problem is known [10]. Some works on the optimality of the BSC model for test channels have been done in [11] and [12]. In these studies, the BSC model is assumed for the test-channel. The BSC model leads to a computable theoretical rate-distortion bound. Moreover, a convergence analysis for the decoding of the binary CEO problem is presented in [13].

The rest of this paper is organized as follows: In section II, the binary CEO problem is introduced and the logarithmic loss are provided. The joint and successive decoding schemes are described in section III. Simulation results and comparisons are presented in section IV. Finally, section V concludes this paper.

## II. The Binary CEO Problem

In this section some preliminaries are presented. First, the system model for the binary CEO problem is illustrated. Next, the concept of the logarithmic loss is given in summary. This distortion measure provides a soft reconstruction for the binary remote source which has some real-world applications [11]. The rate-distortion bound of the two-link binary CEO problem under logarithmic loss is given for the case of the BSC model for the encoders [11]. Some extensions are provided in [12].

*A. System Model*

For the sake of simplicity, we illustrate a two-link binary CEO problem configuration. The scenario can be easily extended to the higher number of links. In real-world applications, there are more than two links but most of the ideas and facts are similar.

A binary symmetric source generates a binary sequence $(x_1, x_2, \ldots, x_n)$ of length $n$ with equal probability 0.5 for 0 and 1. This source is corrupted by two Bernoulli noises $N_1^n$ and $N_2^n$ with parameters $p_1$ and $p_2$, respectively. These noises are mutually independent of the main source $X^n$. The corrupted versions of the main remote source $X^n$ are called $Y_1^n$ and $Y_2^n$. Therefore,

$$Y_1^n = X^n \oplus N_1^n,$$

$$Y_2^n = X^n \oplus N_2^n, \quad (1)$$

where $\oplus$ is the binary addition. In practical applications, the binary sequences $Y_1^n$ and $Y_2^n$ are called noisy observations of the remote source.

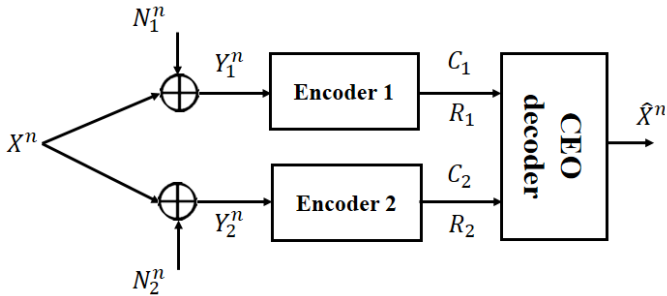

Fig. 1. Two-link binary CEO problem.

Two encoders aim to map the noisy observations to the codewords $C_1$ and $C_2$ from a specified codebook. In this work, we have a source coding approach. Thus, we define the compression rates of the first link and the second link as $R_1$ and $R_2$, respectively. This scenario is completely depicted in Fig. 1.

The point to point source coding theorem implies a theoretical rate-distortion relation $R = 1 - h_b(d)$, where parameters $R$ and $d$ show the compression rate and the distortion, respectively, and

$$h_b(d) = -d \log_2 d - (1-d) \log_2(1-d). \quad (2)$$

In this case, the main source and its reconstruction are binary and the Hamming distance is a suitable criterion for measuring the distortion between the main source and its reconstruction.

Obviously, the Hamming distortion provides a hard reconstruction regardless of the coding scheme. As it is mentioned, the achievable rate-distortion region of the binary CEO problem is unknown under the Hamming loss. In this case, there are inner and outer bounds with a gap [4]. The binary CEO problem is a multi-terminal source coding problem and hence we consider the logarithmic loss as a metric for measuring total distortion.

*B. The Logarithmic Loss Criterion*

The logarithmic loss is a criterion for measuring distortion which provides soft reconstruction and is defined as follows [10].

*Definition:* symbol-wise logarithmic loss between a source symbol $x_j$ and $\hat{x}_j$ is as follows:

$$d(\hat{x}_j, x_j) = \log(\frac{1}{\hat{x}_j(x_j)}), \quad (3)$$

where, $\hat{x}_j(x_j)$ is the soft reconstruction of the main source symbol $x_j$. The total value of this distortion can be calculated by averaging over all symbol-wise terms. Thus,

$$d(\hat{x}^n, x^n) = \frac{1}{n}\sum_{j=1}^{n} d(\hat{x}_j, x_j) = \frac{1}{n}\sum_{j=1}^{n} \log(\frac{1}{\hat{x}_j(x_j)}). \quad (3)$$

It is shown that the logarithmic loss is a universal and optimal criterion for measuring distortion in lossy source coding and compression problems [14-15].

Theoretical rate-distortion bound of the two-link binary CEO problem under the logarithmic loss is obtained in [10] and it is as follows:

$$R_1 \geq I(Y_1; U_1 | U_2, Q),$$

$$R_1 \geq I(Y_2; U_2 | U_1, Q),$$

$$R_1 + R_2 \geq I(Y_1, Y_2; U_1, U_2 | Q),$$

$$D \geq H(X | U_1, U_2, Q). \quad (4)$$

Without loss of generality and optimality, we can consider that the time sharing variable $Q$ is a constant [11]. Furthermore, the cardinality on the alphabet set of the random variables $U_1$ and $U_2$ is 2 [10]. By optimizing the lower bound of the distortion $D$ subject to have a fixed value for the lower bound of the sum-rate $R_1 + R_2$, the optimal test-model for the encoders of links has been obtained [11]. Note that this optimization problem is not convex and it is solved numerically. Based on the numerical results of the Monte Carlo Simulation, the BSC is the optimal model of the encoders as test-channels. Therefore, we model two encoders of the first and second link by two BSCs with crossover probabilities of $d_1$ and $d_2$, respectively.

These models provide a numerically computable inner bound for the rate-distortion region of the binary CEO problem under logarithmic loss. One can find more analytical and numerical results in this regard in [10-12].

## III. DECODING SCHEMES FOR THE BINARY CEO PROBLEM

A two-link binary CEO problem is equivalent to two binary Wyner-Ziv problems. The decoded data in each Wyner-Ziv link can be utilized as a side information at the other link. Hence, different scenarios can be considered for the encoder and decoder designs. The proposed lossy encoders in this paper are based on the compound LDGM-LDPC codes [16-17]. These codes have nested structure and are designed to saturate the binary Wyner-Ziv theoretical bound in each link [17]. A compound LDGM-LDPC code performs a binary quantization by employing an LDGM code at the first step of encoding. This step is accomplished by the Bias-Propagation algorithm [18] and produces binary sequences $U_1^n$ and $U_2^n$. Then in the second step of encoding, it finds the syndromes of $U_1^n$ and $U_2^n$ by using LDPC codes. These syndromes have a length shorter than $n$ and they are sent to the receiver side. In Fig. 2 and Fig. 4, the syndrome generator block is depicted by "SG".

### A. The Joint Decoding Scheme

In the joint decoding scheme, the received syndromes $S_1$ and $S_2$ are jointly decoded. These syndromes are obtained through two steps by using compound LDGM-LDPC codes. In the first step, two noisy observations $Y_1^n$ and $Y_2^n$ are quantized to $U_1^n$ and $U_2^n$ by employing LDGM part of the compound codes. Next in the second step, syndromes of the quantized sequences are computed by using the LDPC part of the compound codes. Finally, these compressed syndromes are sent to the decoder side. The block diagram of the encoder for joint decoding scheme is shown in Fig. 2.

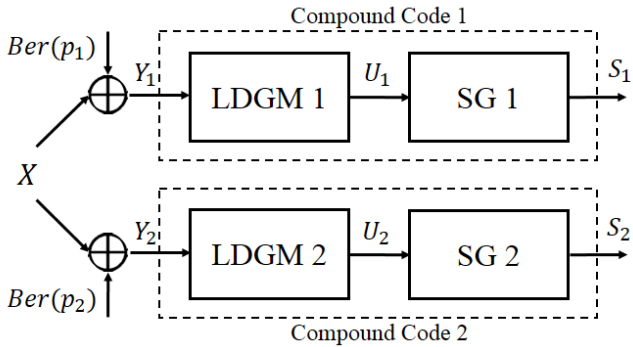

Fig. 2. The encoding scheme for the joint decoding scheme.

At the decoder, $\widehat{U}_1^n$ and $\widehat{U}_2^n$ are decoded with a probability of error. This step is realized by using a Joint Sum-Product (JSP) algorithm, which is an extended version of the Sum-Product (SP) algorithm [19]. Note that, the JSP utilizes the same LDPC codes of the compound codes which are employed at the encoder part. Then, the soft reconstruction of the main source is calculated as follows:

$$\hat{x}_j = \Pr\{x_j | \hat{u}_{1,j}, \hat{u}_{2,j}\}. \qquad (5)$$

The block diagram of the joint decoding scheme is depicted in Fig. 3.

### B. The Successive Decoding Scheme

In the successive decoding scheme, the remote source is decoded in a step by step procedure. In each step, the decoding process is done by employing a side information which comes from previous steps. In this scheme, the error probability of the decoding decreases in each step by inserting more side information. Therefore, smaller BER values is achievable in higher number of links.

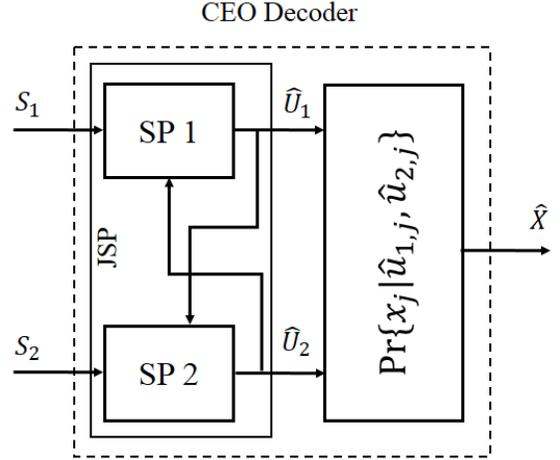

Fig. 3. The joint decoding scheme.

In the encoder part, the first link employs a compound LDGM-LDPC code to generate syndrome $S_1$. In the second link, there is an LDGM code for the binary quantization without any SG block. As it is seen in Fig. 4, there is no symmetry between the links. Thus, one can change the role of the links and achieve to the other corner points in the achievable rate-distortion region. Obviously, the complexity of the joint decoding scheme is more than the successive decoding scheme.

The block diagram of the successive decoding scheme is presented in Fig. 5. As it is shown, there is a SP algorithm at the decoder. The SP algorithm decodes the quantized data in the link that contains the compound structure with the help of a side information coming from the other link. Similar to the joint decoding scheme, there exists a soft decision rule based on the logarithmic loss for reconstructing the remote source. Finally, the soft reconstruction of the main source is calculated as follows:

$$\hat{x}_j = \Pr\{x_j | \hat{u}_{1,j}, u_{2,j}\}. \qquad (6)$$

In both decoding schemes, the compressed data are obtained in a two-step procedure. Similarly at the decoder side, the final soft reconstruction is calculated in a two-step procedure. The empirical distortion in the implementation is calculated by uniform averaging on the logarithmic loss. Note

that (5) and (6) is calculated by using the following Markov chain:

$$U_1 \overset{d_1}{\leftrightarrow} Y_1 \overset{p_1}{\leftrightarrow} X \overset{p_2}{\leftrightarrow} Y_2 \overset{d_2}{\leftrightarrow} U_2, \quad (7)$$

where, the parameters over the arrows indicate the crossover probabilities of the BSC between the variables. Furthermore, the rate values $R_1$ and $R_2$ are computed based on the code length in the compound LDGM-LDPC structures of the first and the second links.

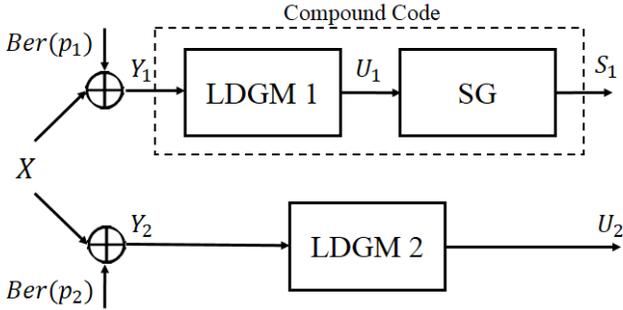

Fig. 4. The encoding scheme for the successive decoding scheme.

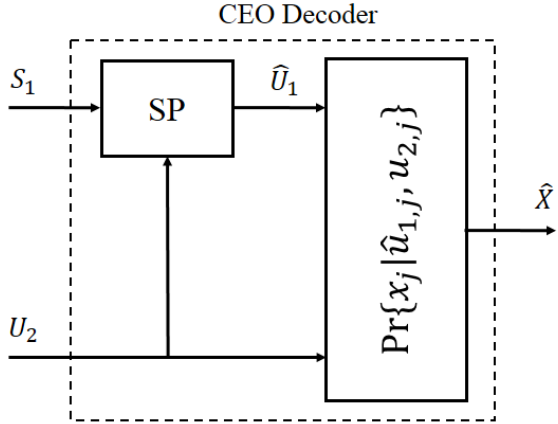

Fig. 5. The successive decoding scheme.

IV. SIMULATION RESULTS AND DISCUSSION

In this section, we present some numerical results around the comparison of the joint and successive decoding schemes from rate-distortion point of view. In a two-link binary CEO configuration, there are two corner points for a fixed value of the sum-rate $R_1 + R_2$. All points on a line segment connecting two end points are intermediate points. Basically, the joint decoding scheme is appropriate for achieving intermediate points of the rate-distortion region [11]. In contrast, the successive decoding scheme aims to achieve corner points of the rate-distortion region [11].

We implement our proposed coding schemes, including the joint and successive decoding schemes, for some intermediate points in achievable rate-distortion region of the two-link binary CEO problem. Then, we determine the total value of the sum-rate and empirical distortion. Finally, we measure the gap from the theoretical bounds to compare the performance of the suggested techniques. Specifically, the successive decoding scheme is more applicable and practical because it has an efficient and low-complexity implementation in higher number of links.

In our implementations, all LDPC codes are designed based on the optimized degree distributions over the BSC which are available in [20]. Furthermore, nested LDGM codes in the compound structure are designed according to the code design method in [17].

By assuming BSC test model for the encoders in each link, two BSC are considered with crossover probabilities $d_1$ and $d_2$. These crossover probabilities yields the following theoretical bounds:

$$\begin{aligned} R_1 &\geq h_b(p*d) - h_b(d_1), \\ R_2 &\geq h_b(p*d) - h_b(d_2), \\ R_1 + R_2 &\geq 1 + h_b(p*d) - h_b(d_1) - h_b(d_2), \\ D &\geq h_b(p_1*d_1) + h_b(p_2*d_2) - h_b(p*d), \end{aligned} \quad (8)$$

where, $p*d = p(1-d) + d(1-p)$, $p = p_1*p_2$, and $d = d_1*d_2$. Note that the concatenation of two BSCs with crossover probabilities $d$ and $p$ yields a BSC with crossover probability of $p*d$. One can find the optimal values of these crossover probabilities shown with $d_1^*$ and $d_2^*$ by solving the optimization problem of the rate-distortion theory which is as follows:

$$\min_{d_1^*, d_2^*} D$$
$$\text{s.t. } R_1 + R_2 = R, \quad (9)$$

for a fixed value of $R$. This optimization problem is not convex, and hence it is solved numerically by using the Lagrangian multiplier method. This problem is completely solved and theoretical bounds of the sum-rate-distortion is derived for a two-link binary CEO problem in [11]. These bounds are a proper criterion for the performance evaluation of different practical coding schemes. Obviously, the coding schemes with smaller gap values from the theoretical bounds is preferable in practical implementations. Note that the complexity is another factor for comparing various coding schemes.

Here, we report some numerical results obtained from computer simulations which compare the sum-rate-distortion performance of the joint decoding scheme and the successive decoding scheme for a two-link binary CEO problem under logarithmic loss.

Parameters in our simulation results are as follows. The noises in the links are with crossover probabilities of $p_1 = 0.15$ and $p_2 = 0.15$. For this case, optimal target distortion values which model the test-channel crossover probabilities have been calculated [11]. For some special sum-rate and distortion values, we calculated optimal target distortion values by solving the optimization problem (9). Our simulations are for the cases of $(d_1^*, d_2^*) = (0.01, 0.01)$,

(0.1,0.1), and (0.1,0.3). The code block lengths are assumed to be $n = 10^5$.

In our simulations, the maximum number of iterations in the Bias-Propagation algorithm is 25. Its value in the SP algorithm is 100. Furthermore, in the JSP algorithm, we have 40 local iterations and 15 global iterations.

For a special value of the sum-rate and distortion, there are infinite number of the intermediate points in the achievable rate-distortion region. We implement the joint and successive decoding schemes to achieve some intermediate points on the theoretical sum-rate-distortion bound. Then, we graphically compare the obtained results based on the gap values from the theoretical bounds.

In Fig. 6, the empirical achieved points from the simulation results are depicted for the mentioned parameters.

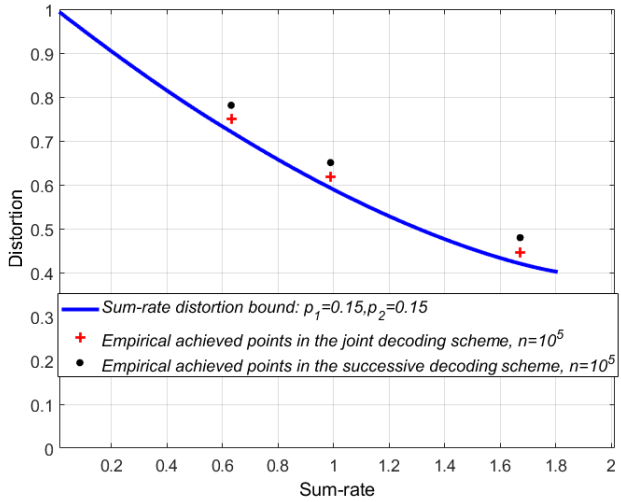

Fig. 6. Sum-rate-distortion performance of the joint and successive decoding schemes.

It is seen from Fig.6 that the gap of the joint decoding scheme is smaller than the gap of the successive decoding scheme. Numerically, there is a gap about 0.02 to 0.03 from the theoretical bound for the case of the joint decoding scheme. This gap value is about 0.055 to 0.06 for the case of the successive decoding scheme. Obviously, complexity of the successive decoding scheme is lower than the joint decoding scheme due to using a single LDGM code in one of the links and using a single SP in the decoder instead of JSP.

## V. Conclusion

In this paper, we compared the performance of the joint and the successive decoding schemes for a two-link case binary CEO problem under logarithmic loss. Encoding and decoding algorithms are based on the compound LDGM-LDPC codes. Under the BSC assumption for the test-channels, a theoretical bound was derived. Based on the simulation results, the joint decoding scheme performs better than the successive decoding scheme in rate-distortion point of view. The work in this paper can be extended to an $L$-link binary CEO problem and multiterminal source coding problem.

Besides, these comparisons can be accomplished for the quadratic Gaussian CEO problem as a future study.